%% file: main.tex
\pdfoutput=1

\RequirePackage{tikz} 

\documentclass[letterpaper, 10 pt, conference]{ieeeconf}  

\IEEEoverridecommandlockouts                              

\usepackage{comment}
\usepackage{amsmath,amssymb,amsfonts}
\usepackage{algorithm}
\usepackage[noend]{algpseudocode}
\usepackage{graphicx}
\usepackage{textcomp}
\usepackage{booktabs}
\usepackage{hyperref}
\usepackage{xspace}
\usepackage{balance}
\usepackage{array}
\usetikzlibrary{arrows,calc,positioning,shapes,fit}
\usepackage[sorting=none,maxbibnames=99]{biblatex}
\usepackage{caption}
\let\labelindent\relax
\usepackage{enumitem}
\usepackage{array}
\usepackage{multirow}

\tikzstyle{block} = [rectangle, draw, text centered, rounded corners, minimum height=2em]
\tikzstyle{eq} = [circle, draw, minimum height=1.7em, inner sep=1pt] 
\makeatletter
\newcommand{\gettikzxy}[3]{%
  \tikz@scan@one@point\pgfutil@firstofone#1\relax
  \edef#2{\the\pgf@x}%
  \edef#3{\the\pgf@y}%
}
\makeatother

\usepackage{biblatex}

\def\BibTeX{{\rm B\kern-.05em{\sc i\kern-.025em b}\kern-.08em
    T\kern-.1667em\lower.7ex\hbox{E}\kern-.125emX}}

\title{A General Formulation for the Teaching Assignment Problem: Computational Analysis Over a Real-World Dataset \\
\author{
    Moa~Johannesson$^{1*}$, Lina~Brink$^{1*}$, Alvin~Combrink$^{1}$, Sabino~Francesco~Roselli$^{1}$, Martin~Fabian$^{1}$
    \thanks{$^*$The authors contributed equally to this work. {\tt\footnotesize \{moajohan, linabri\}@student.chalmers.se} }%
    \thanks{$^{1}$Division of Systems and Control, Department of Electrical Engineering, Chalmers University of Technology, G{\"o}teborg, Sweden {\tt\footnotesize \{combrink, rsabino, fabian\}@chalmers.se} }%
    \thanks{This work was partially supported by the Wallenberg AI, Autonomous Systems and Software Program (WASP) funded by the Knut and Alice Wallenberg Foundation}%
}
}

\input{NewCommands}

\begin{document}
\maketitle
\thispagestyle{empty}
\pagestyle{empty} 
\begin{abstract}
\input{sections/abstract}
\end{abstract}
\section{Introduction}
\label{sec:introduction}
\input{sections/introduction}

\section{Teacher Assignment Problem Model}
\label{sec:method}

\input{sections/method}

\section{Experiments}
\label{sec:results}
\input{sections/results}

\section{Conclusion}
\label{sec:conclusions}
\input{sections/conclusion}

\balance
\printbibliography

\newpage

\clearpage
\onecolumn
\section{Appendix}
\label{sec:plots}
\input{sections/appendix}

\end{document}

%% file: NewCommands.tex
\newcommand{\TA}{TA\xspace}
\newcommand{\TAs}{TAs\xspace}

\newcommand{\TAlen}{s}
\newcommand{\courselen}{c}
\newcommand{\tasklen}{t}

\newcommand{\TtoC}{x}
\newcommand{\TAcourseTask}{y}
\newcommand{\TAtotaltime}{h}
\newcommand{\ctaught}{w}
\newcommand{\assignedTAs}{n}

\newcommand{\takencourses}{\kappa}
\newcommand{\TAreq}{\rho}

\newcommand{\TApref}{\pi}
\newcommand{\TAdisq}{{\xi}}

\newcommand{\tasktime}{\tau}
\newcommand{\alloweddev}{\lambda}
\newcommand{\maxcourses}{\mu}
\newcommand{\maxTAs}{\nu}

\newcommand{\minChunkHours}{\epsilon}

\newcommand{\target}{\Theta}

\newcommand{\totTA}{\mathcal{S}}
\newcommand{\totC}{\mathcal{C}}
\newcommand{\totT}{\mathcal{T}}

\newcommand{\POFT}{\phi}
\newcommand{\FPY}{\theta}
\newcommand{\fullTime}{\vartheta}

\newcommand{\aYear}{\upsilon}

\newcommand{\SwitchHard}{\sigma^H}
\newcommand{\SwitchSoft}{\sigma^S}

\newcommand{\newcourses}{z}

%% file: sections/abstract.tex
The Teacher Assignment Problem is a combinatorial optimization problem that involves assigning teachers to courses while guaranteeing that all courses are covered, teachers do not teach too few or too many hours, teachers do not switch assigned courses too often and possibly teach the courses they favor. Typically the problem is solved manually, a task that requires several hours every year. In this work we present a mathematical formulation for the problem and an experimental evaluation of the model implemented using state-of-the-art SMT, CP, and MILP solvers. The implementations are tested 
over a real-world dataset provided by the Division of Systems and Control at Chalmers University of Technology, and produce teacher assignments with smaller workload deviation, a more even workload distribution among the teachers, and a lower number of switched courses. 

%% file: sections/introduction.tex
Personnel scheduling is a cornerstone of operational efficiency in many service-oriented sectors --- like healthcare and education --- where the primary resource to allocate are people.
The challenge of assigning limited human resources to tasks while satisfying complex 
constraints is a classic combinatorial optimization problem.
In practice, this process is frequently handled manually, which, while allowing for implicit inclusion of nuanced preferences, is often time-consuming, prone to errors, and in many cases results in suboptimal solutions.
Consequently, a significant amount of research has aimed to automate these tasks.

A key challenge in implementing automated 
personnel scheduling 
is the difficulty of a priori formalizing all constraints.
Practitioners often possess tacit knowledge, such as interpersonal dynamics and various personal circumstances, which are essentially unmodeled constraints. 
An ``optimal'' solution that violates these implicit rules is practically useless. 
Therefore, 
a solver should not be viewed merely as a one-shot optimizer, but instead as a decision-support tool within an iterative, human-in-the-loop workflow. 
To support this, 
the solver must allow to easily express and incorporate new constraints, and compute its result quickly so as to enable rapid re-solving during planning sessions.

Within the educational domain, the majority of research has focused on problems such as the \emph{University Course Timetabling} or the \emph{University Examination Timetabling}~\cite{CESCHIA20231}, which are primarily concerned with the temporal allocation of events to time-slots and rooms. 
A distinct and equally critical sub-problem is the \emph{Teacher Assignment Problem} (TAP)~\cite{Carter1998, domenech2016milp, UTA_selection2025}.
Unlike rostering (e.g., Nurse Rostering~\cite{burke2004state}), which focuses on temporal shift coverage and sequencing, the TAP is a resource allocation problem. 
It requires matching teachers with heterogeneous skills to specific courses over a longer horizon (e.g., an academic year), often serving as a precursor to a timetabling phase. 
Despite its practical ubiquity, the TAP has received little attention in the literature compared to its timetabling and rostering counterparts~\cite{UTA_selection2025}.

The broader educational timetabling literature has historically favored metaheuristic solution methods due to the complexity of full university timetabling problems~\cite{oude2019practices}.
However,
though hybrid heuristic/exact methods are often state-of-the-art for large-scale timetabling~\cite{CESCHIA20231, kristiansen2013comprehensive}, exact methods remain highly relevant for specific sub-problems like assignment, where optimality guarantees are valuable for fairness.

Mixed-Integer Linear Programming (MILP) can be regarded as an established ``exact'' baseline, however, comparative studies suggest that other modeling paradigms have much to bring to the table. 
For instance, 
Satisfiability Modulo Theories (SMT) has shown promise in adjacent domains, such as 
Nurse Rostering~\cite{combrink2025comparative}
and Job-Shop Scheduling~\cite{roselli2018smt},
while the relatively younger CP-SAT --- which blends SAT logic with constraint programming --- performs comparatively on certain combinatorial optimization problems, such as Petri net optimization~\cite{lennartson2024optimization} and Minimum Spanning Trees~\cite{montemanni2025solving}.
This suggests that paradigms alternative to MILP, like SMT and CP-SAT, may offer distinct advantages for the combinatorial structure of the TAP.
To our knowledge, however, a rigorous comparison of these paradigms specifically for the TAP remains absent from the literature. 

Our contributions in this work are as follows:
\begin{enumerate}
    \item We formulate a TAP model based on real-world data 
    from the Division of Systems and Control, at Chalmers University of Technology, G{\"o}teborg, Sweden,
    over multiple years, incorporating constraints on workload deviation, competence, and course continuity.
    \item We implement and benchmark this model using solvers based on three distinct paradigms: SMT via Z3~\cite{deMouraBjorner2008Z3}, MILP via Gurobi~\cite{gurobi} and SCIP~\cite{SCIP}, and CP-SAT~\cite{cpsatlp} via OR-Tools. Their computational tractability and suitability for an iterative workflow are evaluated.
    \item We show that our optimization-based approach significantly outperforms historical manual assignment and thereby validate its practical utility.
\end{enumerate}

%% file: sections/method.tex
The teacher assignment problem presented in this work 
concerns
assigning PhD students, henceforth referred to as \emph{Teaching Assistants} or \TAs, to courses taught by the Division of Systems and Control, in the Department of Electrical Engineering at Chalmers University of Technology.
The input data presented in this work is based on real course and staff data provided by the department. 
The following paragraphs introduce the structure and requirements of the courses, and  the \TAs' goals, skills, and preferences.

\paragraph{Teaching Assistant Set} We have a finite set  $\totTA$ of \TAs. Each \TA can teach for up to 5 years, depending on their contract; $\aYear_\TAlen \in \{1,2,3,4,5\}$ is the academic year for \TA $\TAlen \in \totTA$. Moreover, each \TA has a workload $\POFT_\TAlen \in [0,1], \forall \TAlen \in \totTA$ that ranges from $0\%$ to $100\%$, where $100\%$ corresponds to $\fullTime=350$ hours per year. It typically happens for a given year that \TAs do not teach exactly the amount of hours that they are supposed to according to their contract; 
hence, for each \TA 
$\TAlen \in \totTA, \ \FPY_\TAlen \in \mathbb{Z}$
is the amount of teaching hours left from previous years, which can be either positive or negative. From these parameters, the target total workload for this year for each \TA is computed as $\target_\TAlen = \POFT_\TAlen \cdot \fullTime + \FPY_\TAlen$. 

\paragraph{Course and Task Sets} We have a set of courses $\totC$. Each course $\courselen \in \totC$ is characterized by tasks from the set $\totT$, containing \emph{course administration}, \emph{exercise session}, \emph{problem solving session}, \emph{lab session}, \emph{computer session}, \emph{assignment supervision}, \emph{assignment evaluation}, \emph{project supervision}, \emph{exam evaluation}, and \emph{other}. 
Except for \emph{course administration} ($\tasklen^*$ in the model), which is always present, not all courses involve all tasks, and for tasks not present in a course their allotted time is set to $0$.

Based on the number of attending students, each task $\tasklen$ in course $\courselen$ has a required number of \TAs $\TAreq_{c,t} \in \mathbb{N}$ and a total time $\tasktime_{c,t} \in \mathbb{N}$, which is the sum of hours allocated to the \TAs assigned to the task. Finally, in order to avoid \TAs being assigned to tasks for only a very limited time, it is possible to specify the minimum number of hours $\minChunkHours \in \mathbb{N}$ a \TA has to teach a task if they are assigned to it.

\paragraph{Additional Parameters} In addition to the previously defined sets and parameters, some features of the problem regard both courses and \TAs. 
To begin with, \TA $\TAlen$ is able to express their preference $\TApref_{\TAlen,\courselen} \in \{-1,0,1\}$ on course $\courselen$ to say that they are not in favor, indifferent, or in favor of teaching the course. 
$\TAdisq_{\TAlen,\courselen} \in \{0,1\}$ is equal to $1$ if \TA $\TAlen$ is forbidden to teach course $\courselen$, $0$ otherwise. This feature is typically used when \TAs are either not qualified or are not available for a specific period of time due to business or personal reasons. 
Moreover, one of the optimization criteria 
is the consistency of courses assigned to a \TA throughout the years, 
as teaching the same course for multiple years reduces the preparation overhead incurred when a \TA must learn a new subject.
We denote a \emph{new course} as a course that an assigned \TA did not teach in the previous year.
Therefore $\takencourses_{\TAlen,\courselen} \in \{0,1\}$ is equal to $1$ if \TA $\TAlen$ taught course $\courselen$ the year before, $0$ otherwise.

Bounds are given on workload deviation, new courses, number of \TAs per course, and number of courses per \TA.
Since optimization in the SMT formulation is performed by maximizing the number of satisfied soft constraints, there are both hard and soft bounds.
\begin{itemize}
    \item $\alloweddev^H, \alloweddev^S$: hard and soft bounds on workload deviation.
    \item $\SwitchHard, \SwitchSoft$:  hard and soft bounds for the number of new courses.
    \item $\maxcourses^H, \maxcourses^S$: hard and soft bounds for the number of courses per \TA.
    \item $\maxTAs^H$: hard bound on the number of \TAs per course.
    \item $\maxTAs^S$: soft bound for additional \TAs beyond the required number of \TAs for a task $\TAreq_{c,t}$.
\end{itemize}

\subsection{Decision Variables}

The following is the set of variables defined to model the problem:

\begin{itemize}[leftmargin=*]
    \item $\ctaught_{\TAlen,\courselen} \in \{0,1\}$: 1 if \TA $\TAlen$ teaches course $\courselen$, 0 otherwise.
    \item $\TAcourseTask_{\TAlen,\courselen,\tasklen} \in \{0,1\}$: 1 if \TA $\TAlen$ teaches task $\tasklen$ of course $\courselen$, 0 otherwise.
    \item $\TtoC_{\TAlen,\courselen,\tasklen} \in \mathbb{N}$: the number of hours that \TA $\TAlen$ is assigned to task $\tasklen$ in course $\courselen$.
    \item $\TAtotaltime_\TAlen \in \mathbb{N}$: total assigned workload for \TA $\TAlen$.
    \item $\assignedTAs_{\courselen,\tasklen} \in \mathbb{N}$: number of \TAs assigned to task $\tasklen$ in course $\courselen$.
    \item $\newcourses_\TAlen \in \mathbb{N}$: number of new courses $\courselen$ assigned to \TA $\TAlen$.
\end{itemize}

\subsection{SMT Formulation of the Teacher Assignment Problem}
The model includes both hard \eqref{eq:disq}-\eqref{eq:admin} and soft \eqref{eq:soft_target}-\eqref{eq:soft_courses} constraints. Hard constraints must always be satisfied lest the model be infeasible, while soft constraints may be violated without causing the model to be infeasible.

\begin{flalign}
    & \TAdisq_{\TAlen,\courselen} = 1 \rightarrow \sum_{\tasklen \in \totT}\TtoC_{\TAlen,\courselen,\tasklen} = 0 & \forall \TAlen \in \totTA, \courselen \in \totC \label{eq:disq} \\
    & \TAtotaltime_{\TAlen}=\sum_{\courselen \in \totC, \tasklen \in \totT}\TtoC_{\TAlen,\courselen,\tasklen} & \forall \TAlen \in \totTA \label{eq:task_to_total} \\
    & \sum_{\tasklen \in \totT} \TtoC_{\TAlen,\courselen,\tasklen}>0 \rightarrow \ctaught_{\TAlen,\courselen} = 1 & \forall \TAlen \in \totTA, \courselen \in \totC \label{eq:bin_to_int_teacher} \\  
    & \newcourses_\TAlen = \sum_{\courselen \in \totC} \ctaught_{\TAlen,\courselen}(1 - \takencourses_{\TAlen,\courselen}) & \forall \TAlen \in \totTA \label{eq:switch_course}\\
    & \tasktime_{\courselen,\tasklen} = 0 \rightarrow \sum_{\TAlen \in \totTA}\TtoC_{\TAlen,\courselen,\tasklen} = 0  & \forall \courselen \in \totC, \tasklen \in \totT \label{eq:non_task} \\
    & \sum_{\TAlen \in \mathcal{S}} \TtoC_{\TAlen,\courselen,\tasklen} = \tasktime_{\courselen,\tasklen} & \forall \courselen \in \totC, \tasklen \in \totT \label{eq:time_cov} \\
    & \min(\tasktime_{\courselen,\tasklen},\minChunkHours) \leq x_{\TAlen,\courselen,\tasklen} \leq \tasktime_{\courselen,\tasklen} & \forall \TAlen \in \totTA, \courselen \in \totC, \tasklen \in \totT \label{eq:time_bounds} \\
    & \assignedTAs_{c,t} \ge \TAreq_{\courselen,\tasklen} & \forall \courselen \in \totC, \tasklen \in \totT \label{eq:req_staff} \\
    & |\TAtotaltime_{\TAlen}-\target_\TAlen| \leq \lambda^{H} & \forall \TAlen \in \totTA  \label{eq:dev} \\
    & \sum_{c \in \mathcal{C}} \ctaught_{\TAlen,\courselen} \leq \mu^{H} & \forall \TAlen \in \totTA \label{eq:max_course_hard} \\
    & \sum_{s \in \mathcal{S}} \ctaught_{\TAlen,\courselen} \leq \nu^{H} & \forall \courselen \in \totC \label{eq:max_TA_hard} \\
    & \newcourses_\TAlen \leq \SwitchHard & \forall \TAlen \in \totTA \label{eq:hard_switch}\\
    & \assignedTAs_{\courselen,\tasklen^*} \leq 1 & \forall \courselen \in \totC \label{eq:admin}
\end{flalign}
\begin{flalign}
    & \aYear_\TAlen \geq 5 \rightarrow \TAtotaltime_\TAlen = \target_\TAlen & \forall \TAlen \in \totTA \label{eq:soft_target} \\
    & \TAtotaltime_\TAlen \leq \alloweddev^{S} & \forall \TAlen \in \totTA \label{eq:soft_dev} \\
    & \newcourses_\TAlen \leq \SwitchSoft & \forall \TAlen \in \totTA \label{eq:soft_switch} \\
    & \assignedTAs_{\courselen,\tasklen} \leq \TAreq_{\courselen,\tasklen} + \maxTAs^S & \forall \courselen \in \totC, \tasklen \in \totT \label{eq:soft_staff}\\
    & \sum_{\courselen \in \totC} \ctaught_{\TAlen,\courselen} \le \mu^{S}  & \forall \TAlen \in \totTA \label{eq:soft_courses}\\ 
    & \TApref_{\TAlen,\courselen} = 1 \rightarrow \TtoC_{\TAlen,\courselen} = 1 & \forall \TAlen \in \totTA, \courselen \in \totC \label{eq:pref1} \\
    & \TApref_{\TAlen,\courselen} = -1 \rightarrow \TtoC_{\TAlen,\courselen} = 0 & \forall \TAlen \in \totTA, \courselen \in \totC \label{eq:pref2}
\end{flalign}

Constraint~\eqref{eq:disq} forbids \TA-course assignments if the \TA is not available or qualified. 
Constraint~\eqref{eq:task_to_total} connects the number of hours a \TA teaches in each task of each course to their cumulative number of taught hours.
Constraint~\eqref{eq:dev} defines the workload deviation as the difference between theoretical and actual total workload. 
Constraint~\eqref{eq:bin_to_int_teacher} states that if a \TA teaches any amount of hours in a task of a specific course, that \TA is assigned to the course. This constraint, together with~\eqref{eq:switch_course}, is used to keep track of new courses. 
Constraint~\eqref{eq:non_task} models that if a course does not include a task, no \TA is assigned that task in that course. 
Constraint~\eqref{eq:time_cov} states that the cumulative amount of hours \TAs are teaching a task must be equal to the task required time. 
Constraint~\eqref{eq:time_bounds} bounds the number of hours that a \TA is assigned to a task to be larger than $\minChunkHours$ and at most equal to the total time for that task. 
Constraint~\eqref{eq:req_staff} ensures that a task is assigned at least the required number of \TAs for that task. 
Constraint~\eqref{eq:max_course_hard} ensures that each \TA must not teach more than the allowed amount of courses.
Constraint~\eqref{eq:max_TA_hard} ensures that each course must not have more than the allowed amount of \TAs assigned to it. 
Constraint~\eqref{eq:hard_switch} limits the number of new courses a \TA can be assigned.
The task \emph{course administration} may only be assigned to a single \TA, a feature expressed by~\eqref{eq:admin}.

Constraints~\eqref{eq:soft_target}--\eqref{eq:pref2} are \emph{soft}, and while the goal is to satisfy as many of them as possible, violating any number of them shall not result in an infeasible solution. Each soft constraint is associated with a weight, allowing the solver to balance competing objectives.
Constraint~\eqref{eq:soft_target} ensures that if a \TA is in their fifth year, their assigned hours should equal their target hours, so possibly the deviation will be zero.
Constraint~\eqref{eq:soft_dev} sets the assigned hours for a \TA to be below a soft bound.
Assigning new courses is penalized by~\eqref{eq:soft_switch} by softly constraining it to remain below a threshold. 
Constraint~\eqref{eq:soft_staff} softly limits the number of \TAs assigned to a task. 
Constraint,~\eqref{eq:soft_courses} sets the courses taught by each \TA below a given soft bound.
Finally, \TAs' preferences are accounted for in~\eqref{eq:pref1} and~\eqref{eq:pref2}, rewarding assignments with \( \TApref_{\TAlen,\courselen} = 1 \) and penalizing assignments with \( \TApref_{\TAlen,\courselen} = -1 \). 

In order to run the model using the MILP solvers and the constraint programming solver, all the constraints had to be linearized. This involved the definition of additional sets of decision variables, as well as additional constraints to connect the new variables to the existing ones. Moreover, the new model only involves hard constraints and the soft constraints are instead accounted for in the objective function. For the full formulation we refer the reader to the implementation provided on our Github repository\footnote{https://github.com/Chalmers-Control-Automation-Mechatronics/TASP \label{gitlink}}.

%% file: sections/results.tex
\begin{table}[t]
    \centering
    \caption{Parameters and weights for hard and soft constraints respectively for each year.}
    \label{tab:pars}
    \begin{tabular}{ccccccc}
        \toprule
        \textbf{Constraint} & \textbf{Type} & \textbf{2022} & \textbf{2023} & \textbf{2024} & \textbf{2025} & \textbf{2026} \\ 
        \midrule
        \eqref{eq:dev} & hard & 100 & 120 & 150 & 150 & 160 \\ 
        \eqref{eq:max_course_hard} & hard & 2 & 3 & 3 & 2 & 3 \\ 
        \eqref{eq:max_TA_hard} & hard & 8 & 8 & 8 & 8 & 8 \\ 
        \eqref{eq:hard_switch} & hard & 1 & 1 & 1 & 1 & 1 \\ 
        \eqref{eq:soft_dev} & soft & 30 & 30 & 30 & 30 & 30 \\ 
        \eqref{eq:soft_staff} & soft & 1 & 1 & 1 & 1 & 1 \\ 
        \eqref{eq:soft_courses} & soft & 5 & 5 & 5 & 5 & 5 \\ 
        \eqref{eq:soft_switch} & soft & 0 & 0 & 0 & 0 & 0 \\ 
        \bottomrule
    \end{tabular}%
\end{table}

In this section, we present the experimental set up used to evaluate the goodness of our model formulation, as well as the performance of the different solvers. When evaluating the SMT solver Z3, the optimization mode is enhanced, and soft constraints are added with associated weights. The solver searches for solutions that satisfy all hard constraints while minimizing the total penalty induced by soft constraint violations, according to their weights. When evaluating the MILP solvers and the CP solver, the optimization is performed by minimizing an objective function while satisfying the hard constraints. Each term of the objective function represents one of the soft constraints, with a coefficient corresponding to its weight. 

The five problem instances used in the evaluation originate from the real-world instances of assigning PhD students as \TAs to courses, with schedules available for the years 2022 to 2026.
Each instance counts 50 \TAs and 45 courses, each with up to 10 tasks to be assigned. 
No data exists for the \TAs' preferences to courses since this was not taken into consideration. Therefore, all \TA preferences are set to $0$.
The data is anonymized and made available\footref{gitlink}, together with the implementation of the SMT, CP, and MILP models, as well as visualization support functions.


The solvers are tested on the same data as the manual solution each year, to fairly evaluate the goodness of the solution produced. Alternatively, the solver output from year $x$ could be used as input for year $x+1$ to evaluate the benefit of long-ter use of the solver; However, this study is not included in this work. 
The parameter values for hard and soft constraints, shown in Table~\ref{tab:pars}, are fine-tuned for each year and are based on expert advise (the current person responsible for the schedule) and comparisons with the manual scheduling.
The different solvers are compared with manual scheduling on workload deviation, number of assigned courses, and number of new courses per \TA.

\begin{table*}[ht]
\centering
\caption{Time in seconds required by each solver and corresponding Root mean squared error (RMSE) between scheduled and target workloads for solver-based and manual \TA assignments, computed for the five problem instances. When the time is equal to 3600 it means the solver has reached the time limit.}
\begin{tabular}{lrrrrrrrrrr}
\toprule
 & \multicolumn{2}{c}{2022} & \multicolumn{2}{c}{2023} & \multicolumn{2}{c}{2024} & \multicolumn{2}{c}{2025} & \multicolumn{2}{c}{2026} \\
 \midrule
 & \multicolumn{1}{l}{RMSE} & \multicolumn{1}{l}{Time} & \multicolumn{1}{l}{RMSE} & \multicolumn{1}{l}{Time} & \multicolumn{1}{l}{RMSE} & \multicolumn{1}{l}{Time} & \multicolumn{1}{l}{RMSE} & \multicolumn{1}{l}{Time} & \multicolumn{1}{l}{RMSE} & \multicolumn{1}{l}{Time} \\
 \midrule
Z3 & 68.70 & 3600 & 100.95 & 3600 & 128.78 & 3600 & 116.51 & 3600 & 119.94 & 3600 \\
CP-SAT & \textbf{2.84} & 1349 & \textbf{0.22} & 1934 & \textbf{24.82} & 605 & \textbf{38.2} & 3600 & \textbf{47.76} & 49 \\
SCIP & 25.06 & 3600 & 24.81 & 3600 & 35.72 & 3600 & 45.17 & 3127 & 52.84 & 133.53 \\
Gurobi & 22.98 & \textbf{33.21} & 22.57 & \textbf{95.72} & 34.61 & \textbf{16.42} & 44.95 & \textbf{17.29} & 52.84 & \textbf{14.01} \\
Manual & 92.97 & \multicolumn{1}{c}{-} & 95.72 & \multicolumn{1}{c}{-} & 125.59 & \multicolumn{1}{c}{-} & 132.8 & \multicolumn{1}{c}{-} & 144.94 & \multicolumn{1}{c}{-}\\
\bottomrule
\end{tabular}
\label{tab:SolvingTimesAndRMSE}
\end{table*}

We begin by comparing the computation times shown in Table~\ref{tab:SolvingTimesAndRMSE}. For each solver, a time limit of $3600$~seconds is set; 
the solver will return the optimal solution as soon as it is computed and proven to be the optimal, or the best solution found within the time limit.
The SMT solver Z3 timed out on all instances. The open source MILP solver SCIP timed out on 3 instances out of 5, while the CP solver CP-SAT timed out on the 2025 instance. Finally, the commercial solver Gurobi solved all instances in orders of magnitude less than all other solvers, remaining below 100~seconds on all instances.


We now compare the solvers and manual solution on the deviation between assigned hours and target hours.
Table~\ref{tab:SolvingTimesAndRMSE} shows the root mean squared error (RMSE) of the deviation, calculated for each solver and for the manual solution for each of the years 2022--2026.
The SMT solver Z3 shows the worst performance, sometimes under-performing the manual solution. On the other hand, both the MILP solvers Gurobi and SCIP show a significantly better performance than the manual schedule, with Gurobi being slightly better than SCIP on each problem instance. Finally, CP-SAT outperforms both MILP solvers, showing the best performance on each of the five problem instances. Interestingly, the RMSE for year 2022 and 2023 is ten and one hundred times smaller, respectively, compared to the MILP solvers. When it comes to SCIP, this may be due to the fact that the solver timed out and was unable to find the optimal solution. However, when compared to Gurobi, both solvers found the optimal solution (except in 2025), and yet, of the many solutions with the same objective value, CP-SAT consistently chose one that led to a smaller RMSE. 
This shows how parameter tuning may be needed, depending on not only the problem instance but also on the solver.

\begin{figure}[h]
    \centering
    \includegraphics[width=0.49\textwidth]{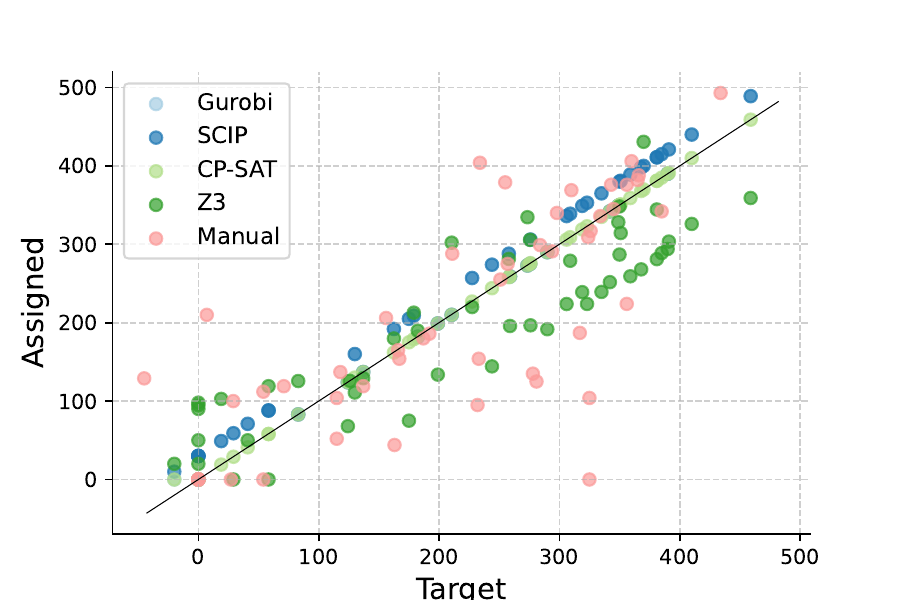}
    \caption{Comparison of assigned versus target hours for each \TA in year 2022 among solvers Gurobi, SCIP, CP-SAT, and Z3, as well as the manual teacher assignment.}
    \label{fig:assVsTarget2022}
\end{figure}

The results of the solvers and the manual schedule on the workload deviation for year 2022 is shown in Figure~\ref{fig:assVsTarget2022}.
Each \TA is represented as a dot with their target hours along the x-axis and assigned hours along the y-axis. 
The diagonal line shows the aim of aligned target and assigned hours. 
Dots above and below the reference line represent \TAs that are over- and under-scheduled, respectively.
It is immediately noticeable how the solutions from CP-SAT almost exactly coincide with the reference line; 
SCIP and Gurobi have almost identical solutions --- close to the reference line but consistently slightly above target; 
Z3 performs significantly worse, with most \TAs teaching below target; 
even so, the result is still visibly better than the manual schedule. Similar plots are produced for the remaining years 2023 to 2026 and available in the appendix in Figure~\ref{fig:AssignedVsTarget_2023-2026}.

\begin{figure}[h]
    \centering
    \includegraphics[width=0.49\textwidth]{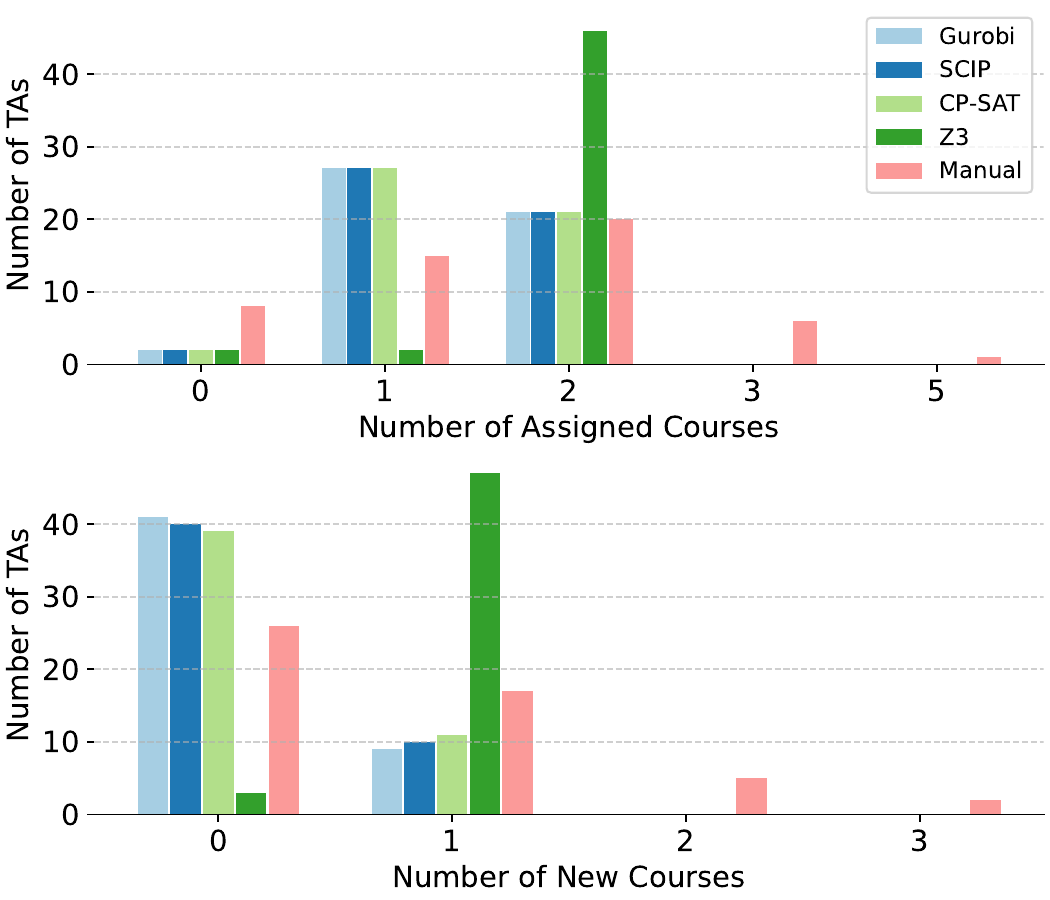}
    \caption{Comparison of the number of assigned courses (upper) and number of new courses (lower) between solvers output and manual teacher assignment for year 2022.}
    \label{fig:assigned2022}
\end{figure}

The solutions produced manually and by the solvers are also compared on the number of assigned courses per \TA and the number of new courses per \TA.
Generally, assigning new courses is undesirable as it involves preparing for the sessions to teach and, possibly, learning new topics; both tasks are time-consuming and typically not accounted for when assigning a new course. 
Therefore we try to keep the number of new courses taught by each \TA as low as possible.
The upper part of Figure~\ref{fig:assigned2022} shows the number of \TAs that are assigned $0$ to $5$ courses, respectively (no student is assigned $4$ courses), 
while the lower part shows the number of \TAs that are assigned $0$ to $3$ new courses, respectively.
The requirements for 2022 were maximum 2 courses per \TA and maximum 1 new course. 
Clearly, the manual schedule violates these constraints, likely due the difficulty of solving such problems manually;
Z3 shows the highest number of \TAs being assigned 2 courses and also the highest number of \TAs being assigned a new course, the vast majority of them. All the other solvers perform identically in terms of number of assigned courses, while performing very similarly in terms of new courses, with Gurobi being slightly better than SCIP, being slightly better than CP-SAT. The difference between Gurobi and CP-SAT confirms our previous statements about parameter tuning needed to achieve identical solutions with different solvers. Similar results are achieved on the instances corresponding to 2023--2026; the reader is referred to figures~\ref{fig:AssignedCourses_2023-2026} and~\ref{fig:SwitchedCourses_2023-2026} in the Appendix. 

%% file: sections/conclusion.tex
In this work we presented a mathematical formulation to model the teacher assignment problem. We implemented the model using state-of-the-art solvers from different communities, both commercial and open source; we tested such solvers on a dataset of real-world instances from the Division of System and Control at Chalmers University of Technology, G{\"o}teborg, Sweden. We optimized the problem in terms of workload deviation from the requirement, workload bounds for the individual \TAs and number of new courses, and we obtained solutions that were superior compared to the manual schedule. Depending on the solver used, we were able to generate a solution within a few seconds up to one hour, while the manual schedule can take several hours over the entire year. 

The obvious next step in this area of research would be to generate synthetic data that mimics the real one and test the scalability of this formulation with the different solvers; it would also be interesting to enrich the data with \TA preferences on courses to evaluate how the corresponding term in the model's objective function affect the solution quality and solving time. Model decomposition could then be applied to the current formulation to increase scalability; finally, an interesting feature that could increase the model's usefulness is to produce and interpret the minimal infeasible constraint set for infeasible instances to provide insight on the cause of infeasibility.

%% file: sections/appendix.tex
\begin{figure*}[!ht]
    \centering
    \includegraphics[width=1\linewidth]{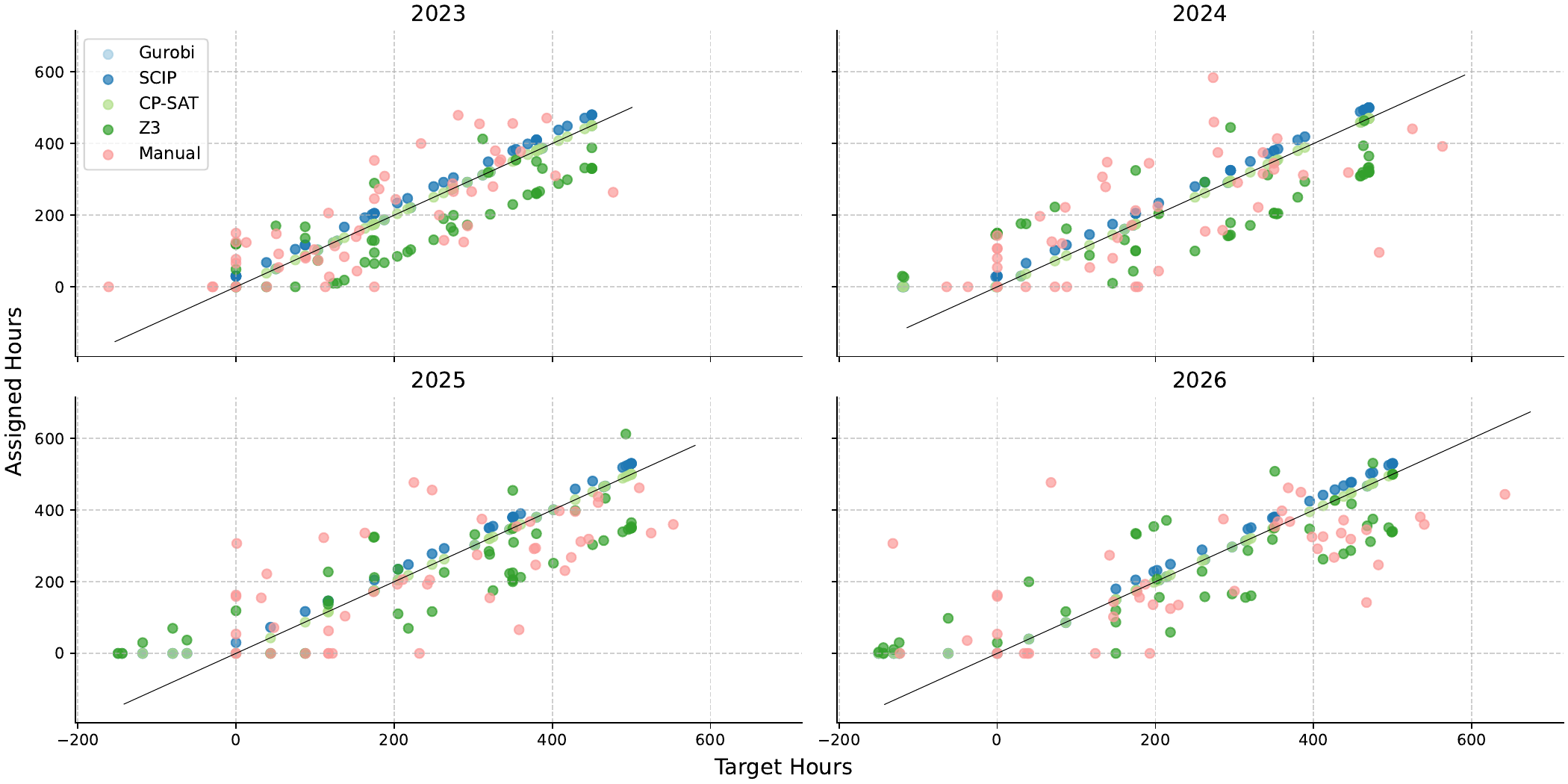}
    \caption{Comparison of the \TAs' assigned versus target hours between solvers and manual assignment for 2023--2026.}
    \label{fig:AssignedVsTarget_2023-2026}
\end{figure*}
\begin{figure*}[!ht]
    \centering
    \includegraphics[width=1\linewidth]{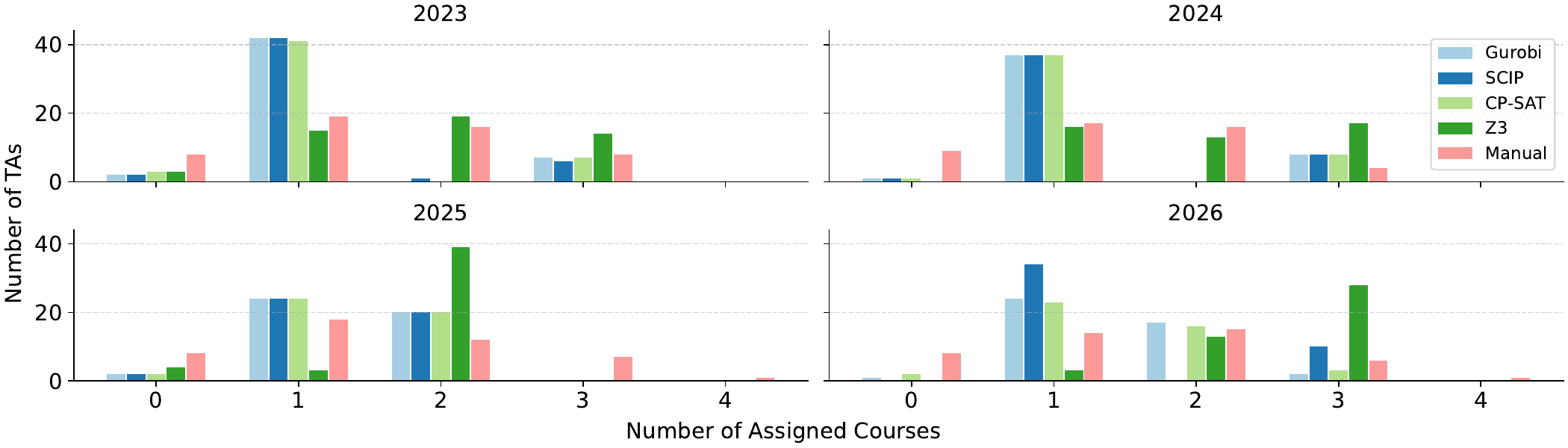}
    \caption{Comparison of the number of assigned courses between solvers and manual assignment for 2023--2026.}
    \label{fig:AssignedCourses_2023-2026}
\end{figure*}
\begin{figure*}[!ht]
    \centering
    \includegraphics[width=1\linewidth]{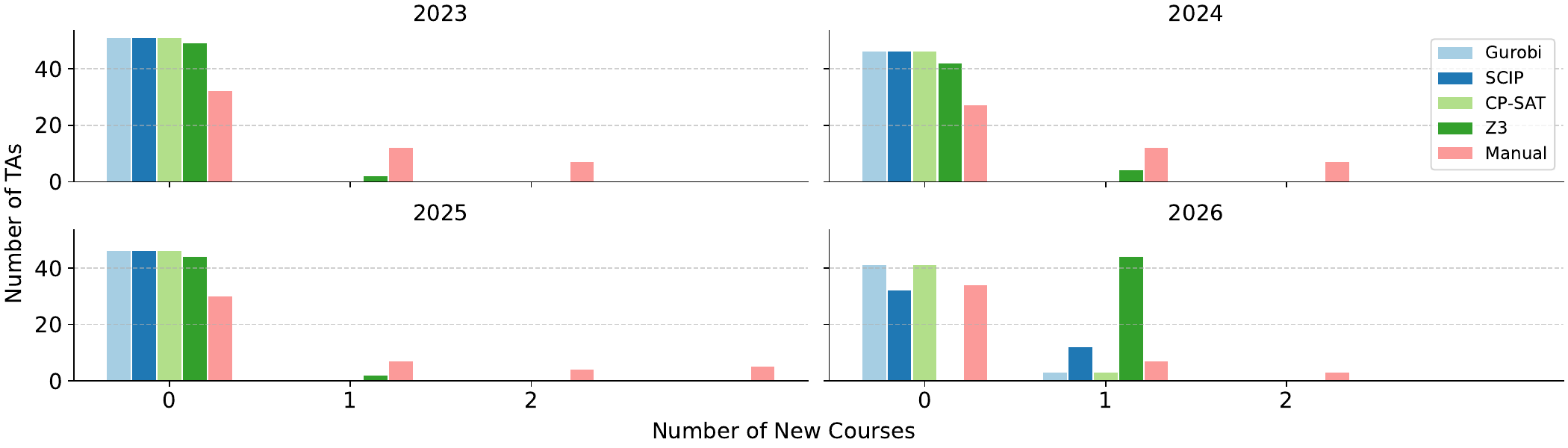}
    \caption{Comparison of the number of assigned new courses between solvers and manual assignment for 2023--2026.}
    \label{fig:SwitchedCourses_2023-2026}
\end{figure*}